\documentstyle[12pt]{article}
\newcommand{\be}{\begin{equation}}
\newcommand{\ee}{\end{equation}}
\newcommand{\ed}{\end{document}}
\newcommand{\lab}[1]{\label{#1}}
\newcommand{\re}[1]{(\ref{#1})}
\newcommand{\ci}[1]{\cite{#1}}
\renewcommand{\baselinestretch}{1.4}
\date{}
\title{ DIFFUSIVE IONIZATION OF RELATIVISTIC HYDROGEN-LIKE ATOM }
\author{ D.U.MATRASULOV \\
 Heat Physics Department of the Uzbek Academy of Sciences,\\
28 Katartal St.,700135 Tashkent, Uzbekistan }
\begin{document}\large
\maketitle

\begin{abstract}
Stochastic ionization of highly excited relativistic hydrogenlike
atom in the monochromatic field is investigated. A theoretical analisis
of chaotic dynamics of the relativistic electron based on Chirikov
criterion is given for the cases of one- and three-dimensional atoms.
Critical value of the external field is evaluated analitically. The diffusion
coefficient and ionization time are calculated.
\end{abstract}

\section*{Introduction}

Inevstigation of properties of the highly excited atoms interacting with
microwave field is important
problem lying at the intersection of several lines of contemporary
research. One of the important themes is chaos. Recently chaotic
dynamics of atoms in the microwave fields have been subject of extensive
research. Much theoretical and experimental work was devoted to this
problem \ci{1,2,3}. A theoretical analisis of behaviour of classical
hydrogen atom, based on the Chirikov criterion \ci{3,4} shows that  for some
critical value of field strength, the electron enters a chaotic regime
of motion, marked by unlimited diffusion leading to ionization.
Up to now much of the discussion on resonance overlap and chaos has
largely been limited by nonrelativistic systems.
It has been known, however, that chaotic dynamics can also been exhibited
by systems undergoing relativistic motion \ci{11}-\ci{14} such as electrons in the free
electron laser and driven relativistic electron plasma wave, driven
relativistic oscillator and relativistic electron in the field of two Kepler
centers.\\
As is well known, \ci{8,9} in the case of hydrogen atom interacting with strong
enough (though still $\epsilon\ll \epsilon_{cr}$) monochromatic wave with
$\Omega \sim \omega_{o} = n^{-3}$ nonlinear oscillations of the atomic electron
become stochastic, that leads to the diffusive ionization of the atom. Diffusion
in the system arises due to the nonlinearity of the classical equations of
motion. First mechanism of diffusive ionization was considered in \ci{Del}.
The phenomenon of diffusive ionization was detected in the numericial
experiments in \ci{Leo}. The difference between diffusive ionization and other
ones is the fact that in this case diffusive excitation corresponds to the
increasing of energy of calassical electron from the initial value to the
ionization one.

In this paper we investigate the transition to chaos in the motion of
a relativistic electron
bound in the Kepler field under the influence of periodic external field.
We generalize the above results on chaotic ionization of the classical
nonrelativistic hydrogen atom to the cases of classical one- and
three-dimensional relativistic hydrogen-like atom.
As is well known (see \ci{8})
and references therein), in the nonrelativistic case one of the remarkable features
of the comparison of the classical theory with the experimental measurements
of microwave ionization is the fact that a one-dimensional model of a hydrogen
atom in an oscillating electric field provides an excellent description of
experimental ionization thresholds for a real three-dimensional hydrogen atom.
In the case of the relativistic hydrogenlike atom, one-dimensional model
enables one to
avoid the difficulties connected with the non-closeness of trajectories in the
relativistic Kepler motion \ci{15} and the resulting additional degrees of freedom
\ci{16}.

Using
Chirikov's criterion for stochasticity, we obtain, in terms of action and charge,
the analytical formula for the critical value of microwave field at which stochastic ionization
will occur.
In this paper we will use the system of units $m_{e} =h= c
=1 $ in which $e^2 = 1/137$.

\section*{One -dimensional model}

Consider the classical relativistic electron in a field of
one-dimensional $-\frac{Z\alpha} {x}$ potential
 where $Z$ is the charge of the center, and $\alpha =
\frac{1}{137}$. The relativistic momentum of this electron is given by $$ p =
 \sqrt{(\varepsilon +\frac{Z\alpha}{x})^2 -1}, $$ where $\varepsilon$ is
the full energy of the electron.  The action is defined as $$ n =
  \int^{x_{1}}_{x_{2}}p dx  = \frac{1}{\pi} a\sqrt{1-\varepsilon^2}, $$ where $ a
 =\frac{\varepsilon Z\alpha}{1-\varepsilon^2}$ $x_1$ and $x_2$ are the
turning points of the electron. From this expression
the Hamiltonian is
\be H_{0} = \frac{n}{\sqrt{n^2 +Z^2\alpha^2}} \lab{unpert} \ee
and the corresponding angular oscillating frequency is
\be
\omega_{0} = \frac{dH_{0}}{dn} =
\frac{Z^2\alpha^2}{(n^2+Z^2\alpha^2)^{\frac{3}{2}}}
\lab{freq}
\ee
It is easy to see that in the nonrelativistic limit (i.e., for small Z)
both Hamiltonian and frequency coincide with corresponding nonrelativistic
ones \ci{1,5}. Now we consider interaction of our atom with monchromatic
field which is written in the form
\be
V(x,t) = \epsilon xcos\omega t ,
\lab{potential}
\ee
where $\epsilon$ and $\omega$ are the field amplitude and frequency.
First we need to write \re{potential} in the action-angle variables. This
can be done by expanding the perturbation \re{potential} into a Fourier series:
\be
V(x,t) = \epsilon \sum^{\infty}_{-\infty} x_{k}(n)cos(k\theta - \omega t),
\lab{Fourier}
\ee
where the Fourier amplitudes of the perturbation are defined by the integral \ci{1,2}
\be
x_{k} = \int^{2\pi}_{0}d\theta e^{im\theta}x(\theta,n) =
-\frac{a}{k}J'(ek) = -\frac{n\sqrt{n^2 +Z^2\alpha^2}}{k}J'(ek),
\ee
$J'_{k}$ is the ordinary Bessel function of order k,
$$e = \frac{\sqrt{n^2 +Z^2\alpha^2}}{n}$$.

Thus the full Hamiltonian of the relativistic hydrogenlike atom in the
monochromotic field can be written in the form
\be
H= \frac{n}{\sqrt{n^2 + Z^2\alpha^2}} +
 \epsilon \sum^{\infty}_{-\infty} x_{k}(n)cos(k\theta - \omega t).
\lab{full}
\ee

For sufficiently small electric fields the Kolmogorov-Arnol'd-Moser
theorem \ci{10} guarantes that most of the straight-line trajectories
in action-angle space will be only slightly distorted by the perturbation.
The maximum distortion of the orbits will occur at resonances where the
phase, $k\theta - \omega t $, is stationary \ci{10}. The resonance
frequency and actions are therefore determined by the relation
\be
k\omega_{0} - \omega = 0
\lab{resonance}
\ee
Then using Eqs. \re{freq} and \re{resonance} the action resonant with the
$k$th subharmonic of the perturbation is
\be
n_{k} = [(\frac{kZ^2\alpha^2}{\omega^2})^{\frac{2}{3}} - Z^2\alpha^2 ]^{\frac{1}{2}}
\lab{action}
\ee

 As is well known there are several methods for investigating the
chaotic dynamics of a system with Hamiltonian in the form \re{full}.
Simplest of them is one, developed by Zaslavsky and Chirikov  \ci{4,6,7}
which is called Chirikov criterion. According to this criterion chaotical
motion occurs when two neighbouring resonances overlap i.e., when the
following condition is obeyed :
\be \frac{\Delta n_{k}}{\delta n_{k}} > 1 ,
\lab{overlap}
\ee
where $\Delta n_{k}$ is the width of $k$ th resonace,
$\delta n_{k} = n_{k+1} - n_{k}$ is the separation of the $k$  and $k+1$
resonances. Using \re{action} this separation is
$$
\delta n_{k} = (\frac{kZ^2\alpha^2}{\omega})^{\frac{2}{3}}\frac{1}{3kn_{k}}
$$
According to \ci{7,10} the width of $k$th resonance defined as
\be
\Delta n_{k} = 4(\frac{\epsilon x_{k}}{\omega'_{0}})^{\frac{1}{2}}
\ee
where
$$ \omega'_{0} = \frac{d\omega_{0}}{dn} = \frac{3n_{k}Z^2\alpha^2}
{(n_{k}^2 +Z^2alpha^2)^{\frac{5}{2}}}. $$
Note that these formulas for $\Delta n_{k}$ and $\delta n_{k}$ in the
limit of small $Z$ coincide for known formulas corresponding to
nonrelativistic case \ci{1}.

Taking into account the expression for $a$ and using the asymptotic formula
$ k^{-1}J'_{k}(ek) \approx 0.411k^{-\frac{5}{3}}$ (for $k>>1$) for the
resonance width we obtain
$$
\Delta n_{k} \approx [\epsilon
k^{-\frac{8}{3}}(Z\alpha)^{-3}(n_{k}^2 +Z^2\alpha^2]^{\frac{1}{2}}$$
Inserting expressions for $\Delta n_{k}$ and $\delta n_{k}$ into
\re{overlap} we have
$$
\epsilon^{\frac{1}{2}}k^{-\frac{1}{3}}(Z\alpha)^{-\frac{3}{2}}n_{k}
(n_{k}^2 +Z^2\alpha^2)^{\frac{1}{2}} >1
$$
This gives us the critical value of the field amplitude at which
sotchastic ionization of the relativistic electron binding in the
field of charge $Z\alpha$ will occur:
$$
\epsilon_{cr} = k^{\frac{2}{3}}(\pi Z\alpha)^3 n_{k}^{-2}(n_{k}^2 +
Z^2\alpha^2)^{-1}
$$
For small $Z$ the last formula can be expanded into series as follow:
$$
\epsilon_{cr} = k^{\frac{2}{3}}(Z\alpha)^3n_{k}^{-4}(1 - \frac{Z^2\alpha^2}{n_{k}^2}
+ ... )
$$
or
\be
\epsilon_{cr} = \epsilon_{nonrel}(1 - \frac{Z^2\alpha^2}{n_{k}^2}
+ ... )
\lab{onedim}
\ee
where  $\epsilon_{nonrel}$ is the critical field corresponding to the
nonrelativistic case.

As seen from this formula the critical field required for stochastic
ionization of relativistic hydrogen-like atom is less than the
corresponding nonrelativistic one.

\section*{Three-dimensional model}

Hamiltonian of the relativistic electron moving in the Coulomb field in terms
of action-angle variables can be written as \ci{16}
\be
H_{0} = [1+ \frac{Z^2\alpha^2}{(n-M+\sqrt{M^2 +Z^2\alpha^2})^2}]^{-\frac{1}{2}},
\ee
where $n = I_{r}+I_{\phi}+I_{\theta}, \;\;\;$ $M=I_{\phi}+I_{\theta},$

$I_{r},I_{\phi},I_{\theta}$ are the radial and angular components of the action.

For
\be
M\gg Z\alpha = \frac{Z}{137}
\lab{cond1}
\ee
this Hamiltonian can be rewritten as
\be
H_{0} = [1+ \frac{Z^2\alpha^2}{(n-M+M)^2}]^{-\frac{1}{2}}\approx ,
\frac{n}{\sqrt{n^2 +Z^2\alpha^2}}.
\ee

The form of this Hamiltonian coincides with the one-dimensional Hamiltonian
obtained in previous section. The frequency is defined by
\be
\omega_{0} = \frac{dH_{0}}{dn} =
\frac{Z^2\alpha^2}{(n^2+Z^2\alpha^2)^{\frac{3}{2}}}
\lab{freq1}
\ee

Consider now (following \ci{9}) interaction of this relativistic three dimensional atom with
external periodic field . Then the full Hamiltonian has the form
\be
H= H_{0}+\epsilon aecos\omega t[-\frac{3}{2}esin\phi +2\sum (x_{k}sin\psi cosk\lambda
+y_{k}cos\psi sink\lambda)],
\lab{ham2}
\ee

where
$$
a = \frac{n}{Z\alpha}\sqrt{n^2+Z^2\alpha^2}
$$
$w$ is the frequency
of external field. To obtain $x_{k}$ and $y_{k}$ in the relativistic case
we consider trajectory equation of a relativistic electron in a Coulomb field \ci{15}:
\be
\frac{p}{r} +ecosq\Phi-1,
\lab{traject}
\ee
where
\be p= \frac{M^2-Z^2\alpha^2}{EZ\alpha}, \;\; q=\sqrt{1-\frac{Z^2\alpha^2}{M^2}},\;\;
e= \frac{\sqrt{E^2M^2-(M^2-Z^2\alpha^2)}}{EZ\alpha},
\lab{exc}
\ee
where $E$ - energy of the electron.

Due to the factor $q$ trajectory of motion of a relativistic electron in the
Coulomb field is not closed \ci{15}. If the conditions \re{cond1} and
$q\approx 1,\;\;\; p\approx \frac{M^2}{E(Z\alpha)},\;\;\;$
are obeyed   $e=\sqrt{1-E^2}$. Therefore eq.\re{traject} can be rewritten as
\be
\frac{p}{r} +ecos\Phi-1,
\ee
Then for the Fourie components one obtains
$$
x_{k} = -\frac{1}{k}J'(ek),\;\;
y_{k}= -\frac{\sqrt{1-e^2}}{ek}J'_{k}(ek),
$$
i.e. in the considered approximation the electron moves along the closed trajectory.
The forms of these expressions for $x_{k}$ and $y_{k}$ coincide with the forms
of nonrelatiivstic ones. But in the relatiivstic expressions $e$ is defined by the
relativistic formula \re{exc}. Let's (following \ci{9}) approximate  Hamiltonian
\re{ham2} with the Hamiltonian of pendulum with mass
$$
M = \mid \frac{d\omega}{dn}\mid^{-1}
$$
Then for the frequency of pendulum one has
$$
\Omega = (\frac{ae}{2M}r_{k}\epsilon_{k})^{\frac{1}{2}},
$$
where $r_{k} = \sqrt{x_{k}^2+y_{k}^2}$

Resonance width can be defined as in \ci{9}
$$
\Delta\nu_{k} = 2\Omega = 2(\frac{1}{2}\omega' aer_{k}\epsilon_{k})^{\frac{1}{2}}
= (2\omega' aer_{k}\epsilon_{k})^{\frac{1}{2}}
$$
The critical value of the external field at which chaotization of motion
of the electron will occur is defined (according to \ci{9})
$$
2.5s^2 > 1,
$$
where
$$
s^2 = \frac{\Delta\nu_{k}+\nu_{k+1}}{\omega_{0}(k)-\omega_{0}(k+1)} =
\frac{(2\omega' r_{k}ae\epsilon)^{\frac{1}{2}} + (2\omega'r_{k+1}ae\epsilon)^{\frac{1}{2}}}
{\omega_{0}(k)-\omega_{0}(k+1)},
$$

$$
{\omega_{0}(k)-\omega_{0}(k+1)} = \frac{\omega}{k}-\frac{\omega}{k+1} =
\frac{\omega}{k(k+1)}
$$

$$
\Delta\nu_{k} = (Z\alpha)^{-\frac{5}{6}}\omega^{\frac{2}{3}}(6e\epsilon)^{\frac{1}{2}}
k^{-\frac{2}{3}}r_{k}^{\frac{1}{2}}[(\frac{\omega}{kZ^2\alpha^2})^{-\frac{2}{3}} -Z^2\alpha^2]^{\frac{1}{2}}
$$

For the critical field we have
\be
\epsilon_{k} = \frac{1}{\gamma_{k}},
\ee

where
\begin{eqnarray}
\displaystyle
\gamma_{k} = 6(Z\alpha)^{-\frac{10}{3}}\omega^{-\frac{2}{3}}ek^{\frac{2}{3}}(k+1)
\{[(\frac{\omega}{kZ^2\alpha^2})^2-Z^2\alpha^2]^{\frac{1}{2}}r_{k}^{\frac{1}{2}}+\\
\nonumber
[(\frac{\omega}{(k+1)Z^2\alpha^2})^2 -Z^2\alpha^2]^{\frac{1}{2}}(1+\frac{1}{k+1})^{-\frac{2}{3}}
r_{k+1}^{\frac{1}{2}}\}
\end{eqnarray}

As is seen from this formula the critical value of the external field
in the relativistic case is less than corresponding nonrelativistic case.

\section*{Diffusive ionization}

As is well known \ci{9}, for the values of the external field strength
exceeding the critical one the motion of the electron becomes chaotic
and its trajectory becomes complicated and unpredicatable. This leads to the
stochastic or diffusive ionization of the atom. According to \ci{9} this process
is of the diffusive character and can be described by the diffusion equation.
Here we calculate coeficient of diffusion following  \ci{9} where it was done for
the nonrelativistic atom. From the Hamiltonian \re{full} one can obtain the
equation for the change of the action $n$:
$$
\frac{dn}{dt} = \frac{\partial H}{\partial\theta} =
\sum F_{k}cos(k\theta-\omega t),
$$
where  $F_{k} = \epsilon kx_{k}$.

Integrating this equation by the same way that was done in \ci{9} we have
$$
\bar{(\Delta n)}^2 = \pi\frac{F_{k}^2}{\omega_{0}}t
$$
From this equation we get the coefficient of diffusion
\be
D = \frac{\pi}{2}\frac{F_{k}^2}{\omega_{0}}
\lab{dif}
\ee
Taking into account expressions for $F_{k}$  and $\omega_{0}$  we have
$$
D = \frac{\pi}{2}\epsilon^2(Z\alpha)^{-6}\omega^2(n^2+Z^2\alpha^2)^{\frac{9}{2}}x_{k}^2,
$$

where
$x_{k} = n\sqrt{n^2 +Z^2\alpha^2}k^{-1}J'_{k}(\xi k),\;\;\;$ $\omega = k\omega_{0}$

For $k\gg1$  coefficient of diffusion can be written as
$$
D \approx \frac{1}{4}\epsilon^2(Z\alpha)^{\frac{2}{3}}\omega^{-\frac{4}{3}}n^2(n^2+Z^2\alpha^2)^{\frac{1}{2}}
$$
For $k=1$
$$
D = \frac{\pi}{2}\epsilon^2(Z\alpha)^{-2}n^2(n^2+Z^2\alpha^2)^{\frac{5}{2}}
$$

In the general case calculation of diffusion coefficient can be performed
analogously.
The value of $D$ can be obtained from \re{dif} by the substitution
$$
x_{k}\longrightarrow \frac{1}{2}(x_{k}^2 + y_{k}^2)e
$$

As is seen from these formulas the coefficient of diffusion in the case
of relativistic atom is greater than for the corresponding nonrelativistic one.
Hence the time
$$\tau_{D}\approx\frac{n^2}{2D},$$

during which ionization of the relativistic atom will occur is less than the
ionization time of the nonrelativistic atom.
The ionization rate ca be defined as
$$
\omega_{D} = \tau_{D}^{-1}
$$
For $k=1$ we have
$$
\omega_{D} = \frac{1}{3}(n^2+Z^2\alpha^2)^{5/2} =
\frac{1}{3}\frac{\epsilon^2}{Z^2\alpha^2}n^5(1+\frac{5}{2}\frac{Z^2\alpha^2}{n^2}+...),
$$
\be
\omega_{nonrel}(1++\frac{5}{2}\frac{Z^2\alpha^2}{n^2}+...),
\lab{rate}
\ee
where
$$
\omega_{nonrel} = \frac{1}{3}\frac{\epsilon^2}{Z^2\alpha^2}n^5
$$
is the ionization rate for the nonrelativistic atom.
Comparing eqs. \re{onedim} and \re{rate} one can see that relativistic corrections 
to the ionization rate are considerable than relativistic corrections to the 
critical field. 

\section*{Conclusion}

We have obtained approximate analytical formula for the critical value
of the field requiring for stochastic ionization of relativistic electron
binding in the Coulomb field of charge $Z$ in terms of $Z$ and action
$n_{k}$. Since the relativistic Rydberg atom is an esentially quantum object,
the study of its microwave field excitation, provides, therefore a
testing ground for the existence of quantum relativistic "chaotic"
phenomena. More detail analysis of the above problem should be
given by comparing the solution of time-dependent Dirac equation with classical equations
of motion. Finally, one should note another problem which is closely related
to the considered above one. Presently there is a considerable interest to
the so-called chaotical autoionization of atoms and molecules \ci{17,18,19}.
The methods of investigations of such a systems are the same as ones
for the difusive ionization \ci{17,18}. Therefore the above results can be
also applied for the theoretical investigation of the autoionization
processes of the relativistic atoms and molecules.

\ed